# Grain Growth with Size-Dependent or Statistically Distributed Mobility

Yanhao Dong and I-Wei Chen*

Department of Materials Science and Engineering, University of Pennsylvania, Philadelphia, PA

19104, USA

**Abstract**

Conventional grain growth is rate-limited by the mobility of grain boundary. To describe similar phenomena limited by the mobility of other grain junctions, we have developed a general theory allowing for size-dependent mobility and its statistical variance. We obtained analytic solutions for the steady-state size distribution and the growth exponent, defined as (grain size)$^n$ ~ time, down to $n=1$, which arises when the mobility of three-grain lines is rate-limiting. When the mobility of four-grain junctions is rate-limiting, exponential growth and a bifurcating size distribution result. These solutions manifest a general trend: The size distribution narrows with increasing $n$. Yet experimentally the opposite trend has been observed recently, which can only be reproduced in simulation if the mobility distribution is made at lease bimodal, with one mode being immobile or nearly immobile. The latter can be realized in slow grain growth below the temperature of mobility transition.

*Corresponding Author Information

**Tel: +1-215-898-5163; Fax: +1-215-573-2128**

**E-mail address: iweichen@seas.upenn.edu (I-Wei Chen)**

**Postal address:** Department of Materials Science and Engineering, University of Pennsylvania, LRSM Building, Room 424, 3231 Walnut St., Philadelphia, PA 19104-6272

I. Introduction

Theory of microstructure coarsening to minimize the total interfacial energy was first formulated by Lifshitz, Slyozov [1] and Wagner [2] (LSW) for precipitates. This capillary process dictates a critical particle size, above which the precipitates grow and below which the precipitates shrink, in what is commonly referred to as the Oswald ripening process. Grain growth in a polycrystal was similarly analyzed by Hillert [3]. As the capillary driving force decreases with the feature size, the coarsening rate characteristically decreases with time. The LSW theory predicts a cubic growth law for bulk diffusion-controlled precipitate coarsening, which has been verified experimentally. Hillert's prediction—of parabolic law for normal grain growth controlled by boundary diffusion—is more difficult to verify because of crystallographic texture and substructure pinning, by sub-grain walls, pores and second phases. However, there is no question of its validity since in high-density ceramics, of few pores and little residual stress, parabolic grain growth has been convincingly demonstrated many times [4-7].

In Hillert's theory, the growth rate (i.e., the equation of motion) of individual grain size $G$ under a capillary driving force $2\gamma/G$ with $\gamma$ being the interfacial energy is written as

$$\frac{dG}{dt} = 2M_b\gamma\left(\frac{1}{G_{cr}} - \frac{1}{G}\right) \qquad (1)$$

Here, $M_b$ is the mobility of grain boundary, and $G_{cr}$ is the critical grain size that neither grows nor shrinks at time $t$ thus setting up a chemical potential $2\gamma/G_{cr}$ that ensures mass (volume) conservation. The theory is incomplete, however, because grain growth in a polycrystal also requires the motion of the entire grain boundary network, which includes not only 2-grain boundaries but also lower dimensional features: 3-grain lines and 4-grain junctions. If the latter are less mobile than 2-grain boundaries, then they can pin the network thus suppress grain growth [8-10]. This effect becomes more severe as the grain size decreases, which is accompanied by a higher concentration of 3-grain lines and 4-grain junctions.

We hypothesize that the grain velocity limited by the mobility of 3-grain line, $M_t$, may be described by

$$\frac{dG}{dt} = 2M_t\gamma\left(\frac{G}{a}\right)\left(\frac{1}{G_{cr}} - \frac{1}{G}\right) \qquad (2)$$

Here, we assume the driving force on a grain boundary of an area $G^2$ is entirely spent on a 3-grain line, which has an effective area of $aG$ with $a$ taken as the atomic spacing. Likewise, we hypothesize the grain velocity limited by the mobility of 4-grain junction, $M_j$, is

$$\frac{dG}{dt} = 2M_j\gamma\left(\frac{G}{a}\right)^2\left(\frac{1}{G_{cr}} - \frac{1}{G}\right) \qquad (3)$$

Here, we assume the entire driving force is spent on a 4-grain junction with an effective area of $a^2$. In analogy with Eq. (1), we can now identify a size-dependent effective mobility $M\left(\frac{G}{a}\right)^\alpha$, where $\alpha$ varies from 0 to 2 when the control feature changes from 2-grain boundaries to 4-grain junctions. Indeed, precipitate coarsening corresponds to $\alpha=-1$.

In the context of mean-field theory, one can immediately obtain the growth law by dimensional analysis, assuming $(1/G_{cr}-1/G)$ is of the order of $1/G$. This gives $(G/a)^{2-\alpha} \sim M\gamma t/a^2$, where $M$ is $M_b$ for $\alpha=0$ in the parabolic law, and $M_j$ for $\alpha=1$ in the linear law. (For precipitate growth, $\alpha=-1$, so it gives the cubic law.) The case of $\alpha=2$ is degenerate, and it leads to the exponential law, or $\ln(G/G_0) \sim M\gamma t/a^2$, where $G_0$ is a reference grain size. Of course, the dimensional analysis cannot provide the proportionality constants in the above growth laws, nor can it provide the size distributions. They require a more detailed analysis, which is provided below for both integer and non-integer $\alpha$ following the analytical method of LSW (for $\alpha=-1$) [1,2] and Hillert (for $\alpha=0$) [3] and verified by numerical simulations.

Our analysis will further include the possibility of inhomogeneous mobility. One obvious extension is the case of mixed control with more than one mobility at play, which as already mentioned is relevant to a polycrystal. Another interesting case entails a bimodal distribution of mobility. The origin of bimodal mobility, or mobility inhomogeneity in general, may come from accumulation of solutes, pores and second-phase particles on the grain boundaries and their junctions, which is a distinct possibility as the grain size increases and the boundary areas/junctions are eliminated. This evolution may also be accompanied by an evolution of grain boundary structure, which relaxes and adopts new configurations. Indeed, inasmuch as grain boundaries are not structureless and structural multiplicity is myriad, statistical variation in grain boundary and junction mobilities is entirely plausible [11,12]. It will become clear in the following analysis that these inhomogeneities impart qualitatively new features that are most relevant to understanding experimental observations during low-temperature growth [13].

## II. Growth Kinetics with Size-Dependent Mobility

(1) Analytical solution by the Lifshitz-Slyozov-Hillert method [1,3]

We start with the generalized mean-field equation of motion for individual grain size $G$

driven by a capillary pressure $2\gamma/G$

$$\frac{dG}{dt} = 2M\gamma\left(\frac{G}{a}\right)^{\alpha}\left(\frac{1}{G_{cr}} - \frac{1}{G}\right) \quad (4)$$

Introducing a relative size $u = \dfrac{G}{G_{cr}}$ in which both $G$ and $G_{cr}$ are time dependent, we first evaluate $\dfrac{du^{2-\alpha}}{dt}$ using Eq. (4). (We could also evaluate $\dfrac{du^s}{dt}$ with any $s$ but it turns out $s=2-\alpha$ is the best choice unless $\alpha=2$, for which we will let $s=0$ as shown later.)

$$\frac{du^{2-\alpha}}{dt} = (2-\alpha)\frac{1}{G_{cr}}\left[\frac{M\gamma}{a^{\alpha}}G_{cr}^{\alpha-1}(u-1) - u^{2-\alpha}\frac{dG_{cr}}{dt}\right] \quad (5)$$

Dividing both sides by $\dfrac{dG_{cr}}{dt}$, we find

$$\frac{du^{2-\alpha}}{dG_{cr}} = (2-\alpha)\frac{1}{G_{cr}}\left[\frac{2M\gamma}{a^{\alpha}}G_{cr}^{\alpha-1}\frac{dt}{dG_{cr}}(u-1) - u^{2-\alpha}\right] \quad (6)$$

Eq. (6) can be rewritten as

$$\frac{du^{2-\alpha}}{d\left(\ln G_{cr}^{*}\right)^{2-\alpha}} = (2-\alpha)\frac{2M\gamma}{a^{\alpha}}\frac{dt}{dG_{cr}^{2-\alpha}}(u-1) - u^{2-\alpha} \quad (7)$$

or

$$\frac{du^{2-\alpha}}{d\tau} = A(u-1) - u^{2-\alpha} \quad (8)$$

by (a) letting $A = (2-\alpha)\dfrac{2M\gamma}{a^{\alpha}}\dfrac{dt}{dG_{cr}^{2-\alpha}}$ be the "growth law", (b) defining a dimensionless grain size $G_{cr}^{*} = \dfrac{G_{cr}}{G_{cr0}}$ where $G_{cr0}$ is the initial $G_{cr}$, and (c) defining a dimensionless time $\tau = \ln G_{cr}^{*}$, which is justified because both $G_{cr}$ and $\ln G_{cr}^{*}$ are expected to monotonically grow with time.

Next, we will ascertain the existence of a steady-state solution, which obtains when $\tau \to \infty$. For this purpose, the choice of $G_{cr0}$ is immaterial since it only affects the choice of $\tau=0$. At $\tau \to \infty$, $\dfrac{dA}{dt}$ is neither positive nor negative or else $A$ will diverge to positive or negative infinity, which is non-physical for a finite-sized sample. So, $\dfrac{dA}{dt} = 0$ and $A$ should be a constant. This leads to a steady-state ($\tau \to \infty$) growth law in the form of $G_{cr}^{2-\alpha} \sim t$ with a growth exponent $n=2-\alpha$. Positive growth of $G_{cr}$ is thus ensured by $A > 0$ for $\alpha < 2$. For $\alpha > 2$, we must have $A<0$ to ensure $1/G_{cr}$ decreases with $t$, but at $\tau \to \infty$ it also gives $1/G_{cr}^{\alpha-2} \sim -t$, which is non-physical. So there is no steady state growth law for $G_{cr}$. Lastly, there is no restriction on how negative $\alpha$ can be to obtain steady state growth.

Following Lifshitz, Slyozov and Hillert, we next show that a steady state solution must make the curve $\dfrac{du^{2-\alpha}}{d\tau}$ tangent to the $u$-axis at a point $u_0$ where it satisfies the following double-root condition [1,3]

$$\left.\dfrac{du^{2-\alpha}}{d\tau}\right|_{u=u_0} = \dfrac{d}{du}\left(\dfrac{du^{2-\alpha}}{d\tau}\right)\bigg|_{u=u_0} = 0 \qquad (9)$$

The condition may be understood by referring to **Fig. 1** at $\tau \to \infty$, where the double-root condition is equivalent to setting $A$ (a constant at $\tau \to \infty$) at a critical value $A_0$. If $A<A_0$, then $\dfrac{du^{2-\alpha}}{d\tau}$ lies below the $u$-axis, so all the grains will shrink and the sample will vanish, which is impossible. If $A>A_0$, then $\dfrac{du^{2-\alpha}}{d\tau}$ intersects the $u$-axis twice at $u_1$ and $u_2$. So all grains between $u_1$ and $u_2$ will converge to $u_2$, which provides a finite population of grains that will grow to infinity at $\tau \to \infty$, thus violating volume conservation. The only allowed case is when $A=A_0$. Importantly, since grains larger than $u_0$ will shrink to $u_0$ but never cross it, there cannot be any finite population of such grains or else their volume will again diverge over time. Therefore, $u_0$ is the upper limit of the grain size at the steady state.

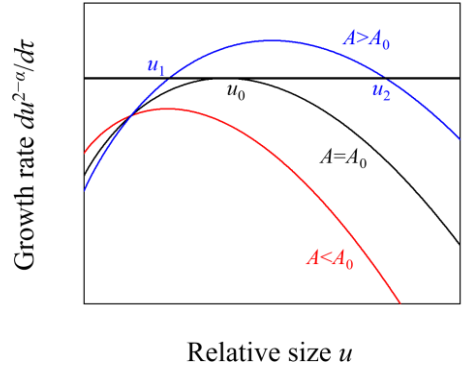

**Figure 1** Schematics showing growth rate $\dfrac{du^{2-\alpha}}{d\tau}$ as a function of relative grain size $u$ with three values of $A$.

By inspection of Eq. (8) and noting $A>0$, we can identify the condition for $\dfrac{du^{2-\alpha}}{d\tau} \to -\infty$ at $u \to \infty$ and $\tau \to \infty$ as

Case I: $\alpha < 1$. Here, the double root condition is satisfied by

$$u_0 = \frac{2-\alpha}{1-\alpha} \qquad (10)$$

$$A = \frac{(2-\alpha)^{2-\alpha}}{(1-\alpha)^{1-\alpha}} \qquad (11)$$

With this $A$, the steady-state growth law becomes

$$\frac{dG_{cr}^{2-\alpha}}{dt} = \frac{2M\gamma}{a^{\alpha}}\left(\frac{1-\alpha}{2-\alpha}\right)^{1-\alpha} \qquad (12)$$

Case II: $\alpha = 1$ and $A=1$ giving $\dfrac{du}{d\tau} = -1$. Here, $u_0$ is at infinity so there is no upper cut-off of grain size at the steady state: All the grains shrink relative to $G_{cr}$ at the steady state regardless of size. With this $A$, the steady-state growth law becomes

$$\frac{dG_{cr}}{dt} = \frac{2M\gamma}{a} \qquad (13)$$

For $1<\alpha<2$, the second term on the right-hand side of Eq. (8) no longer dominates at $u \to \infty$. However, since positive growth of $G_{cr}$ still demands $A>0$, we have the situation of **Fig. 1** with only one root at $u_2$ instead. This leads to bifurcation: All the grains smaller than $u_2$ will shrink, and all the grains larger than $u_2$ will grow indefinitely, which will consume more volume over time. Moreover, since there is not a unique solution for $A$, the above procedure does not lead to a steady-state solution.

Lastly, when $\alpha=2$, we cannot use the procedure starting with Eq. (5). But choosing $s=1$ leads to

$$\frac{du}{d\tau} = u\left[A'(u-1)-1\right] \quad (14)$$

where $A' = \frac{2M\gamma}{a^2}\frac{dt}{d\ln G_{cr}^*}$ gives the exponential growth law mentioned in the **Introduction**. Again, the only possible solution is $A'>0$, but there are now two roots at $u_1=0$ and $u_2=1+1/A'$. All the grains smaller than $u_2$ will shrink to 0 but not disappear since $dG/dt$ vanishes at $G=0$ (which is a general characteristic for $\alpha>1$), and all the grains larger than $u_2$ will grow indefinitely, which will consume more volume over time. Like above, there is not a unique solution for $A'$, and the above result simply rewrites $G_{cr}$ into a form like a Laplace transform, $G_{cr} = \int_0^\infty f(A')\exp\left(\frac{2M\gamma}{a^2}\frac{t}{A'}\right)dA'$, whose solution depends on the initial condition. Obviously, there is no steady state in this case either.

For Case (I-II), we have found their steady-state size distributions, given in the Appendix, following the method of Lifshitz, Slyozov [1] and Hillert [3]. The normalized steady-state grain size distribution function $P(u)$, is defined as

$$P(u) = \frac{\varphi(u,\tau)}{N(\tau)}$$
$$= \frac{\beta}{2-\alpha} \exp\left[-\frac{\beta}{2-\alpha} \int_0^u \frac{-(2-\alpha)u^{1-\alpha}}{A(u-1)-u^{2-\alpha}} du\right] \frac{(2-\alpha)u^{1-\alpha}}{u^{2-\alpha}-A(u-1)} \qquad (15)$$

Where $\beta=2$ in two dimensions and $\beta=3$ in three dimensions. To obtain the average grain size $u_{avg}$ and $G_{avg}$, we use

$$u_{avg} = \int_0^{u_0} uP(u)\,du \qquad (16)$$

$$G_{avg} = u_{avg} G_{cr} \qquad (17)$$

Therefore, there exists a one-to-one relationship between $u_{avg}$ and $\alpha$, and $G_{avg}$ are related to $G_{cr}$ by $u_{avg}$. It follows from Eq. (12) that $\dfrac{dG_{avg}^{2-\alpha}}{dt}$ is also constant at the steady state. That is, the experimentally measured $G_{avg}$ should obey a growth law with the same exponent $n=2-\alpha$: $n=3$ for Oswald ripening at $\alpha=-1$, $n=2$ for normal growth at $\alpha=0$, and $n=1$ for triple-line controlled growth at $\alpha=1$.

Several features of these solutions are noted below. First, although there is no upper cut-off $G_{max}$ for $\alpha=1$, there is one for $\alpha<1$. If normalized with respect to $G_{cr}$, it gives $u_0$ as shown by the blue curve in **Fig. 2**, and if normalized with respect to $G_{avg}$, it gives the red curve in the same figure. Both $u_0$ and $G_{max}/G_{avg}$ increase with $\alpha$ and go to infinity at $\alpha=1$. In contrast, $u_{avg}$, which is the inverse of $G_{max}/G_{avg}$ and shown as the black curve in **Fig. 2**, decreases with $\alpha$ and reaches a minimum of 1/3 at $\alpha=1$. Second, as shown in **Fig. 3a**, $P'(u)$ becomes more extended as $\alpha$ increases, and a similar trend is apparent in the normalized steady-state grain size distribution function $P(G/G_{avg})$ shown in **Fig. 3b**. To quantify the dispersion, shown in **Fig. 4** is the standard deviations of the grain size distribution, $\sigma'$, for relative grain size $u$, which reaches a maximum around $\alpha=0.75$, but the standard deviation $\sigma$ for $G/G_{avg}$ monotonically increases with $\alpha$, reaching a maximum of 1.0 at the steady-state limit, $\alpha=1$. Therefore, as $\alpha$ decreases and $n$ increases, $\sigma$ decreases and a more homogenous size distribution results. This is because a larger $n$ implies a slower growth rate for the larger grains, which in turn allows smaller ones to catch up. Meanwhile, much smaller grains will rapidly shrink out of existence (unless $\alpha>1$ which is not solved above.) In this way, a more narrowly distributed size distribution will result.

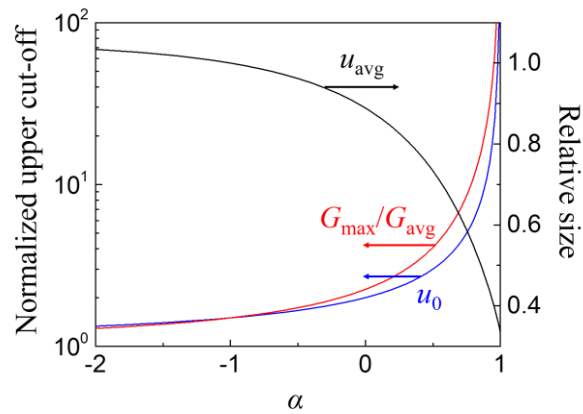

**Figure 2** Calculated normalized upper cut-off $u_0$, $G_{max}/G_{avg}$ and the average relative grain size $u_{avg}$ as a function of $\alpha$.

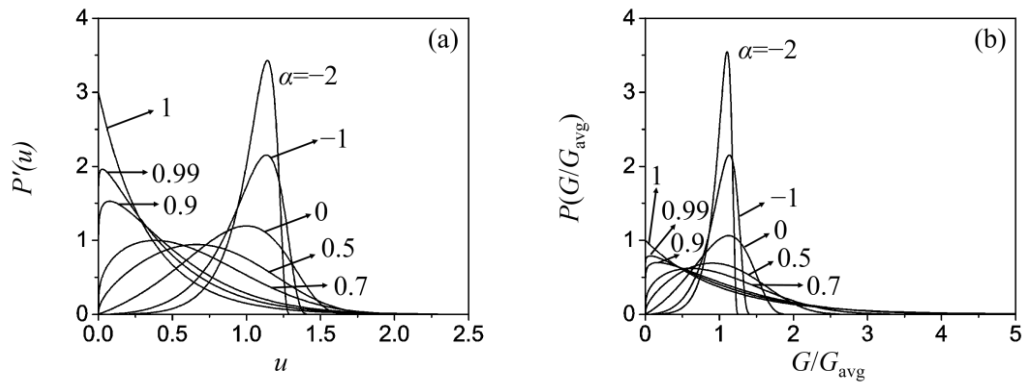

**Figure 3** Calculated normalized steady-state grain size distribution function (a) $P'(u)$ and (b) $P(G/G_{avg})$ as a function of $\alpha$.

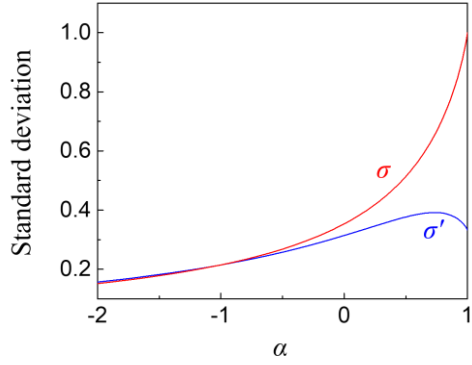

**Figure 4** Calculated standard deviations $\sigma'$ for relative grain size $u$ and $\sigma$ for relative grain size $G/G_{avg}$ as a function of $\alpha$.

(2) Numerical verifications

We conducted numerical simulations [14] to verify the above solution and to explore the cases where a steady-state solution does not exist. Starting with an initial size distribution, the grain size $G$ for each grain is numerically updated according to the equation of motion after a small time-interval. To obtain statistically meaningful results, we typically started with a population of over 1,000,000 grains with the predicted steady-state distribution in the analytic solution, and ended with over 10,000 grains that may be again used to determine the steady-state solution. If the starting grain size distribution is not the predicted steady-state one, our simulations still led to the steady state distribution eventually, but it took a much longer time to converge to the correct slope of the growth kinetic, $\dfrac{dG_{avg}^{2-\alpha}}{dt}$, as have been reported before by Chen, Devenport and Wang [14]. At each time step, mass (volume) conservation is used to update the chemical potential, which is embodied in $G_{cr}$—the only free parameter in the equation of motion that is now constrained by

$$\frac{d\left(\sum G^3\right)}{dt} = \sum G^2 \frac{dG}{dt} = 0 \quad (18)$$

Substituting the equation of motion into the above, we obtain

$$G_{cr} = \frac{\sum G^{\alpha+2}}{\sum G^{\alpha+1}} \quad (19)$$

Starting from the corresponding steady-state size distribution available for cubic, parabolic and linear growth laws in the **Appendix**, we readily verified in our simulation that they remain invariant from $t=0$ and are the indeed steady states (data not shown). For $\alpha=-1$, the simulated $G_{avg}^3(t)$ follows a straight line with a slope 0.445, vs. 4/9 as predicted. We also verified $G_{cr} = G_{avg}$, and the size distribution has a standard deviation $\sigma$ of 0.215 ($\sigma'=0.215$), which remains unchanged over $t$. Similarly, for $\alpha=0$, the simulated $G_{avg}^2(t)$ is a straight line with a slope 0.395, vs. 32/91 as predicted. We also verified $G_{cr} = 1.125 G_{avg}$, and the size distribution has a standard deviation $\sigma$ of 0.354 ($\sigma'=0.314$), which remains unchanged over $t$. For $\alpha=1$, the simulated $G_{avg}(t)$ is shown in **Fig. 5** where we set $\frac{2\gamma M_t}{a} = 1$; it is a straight line with a slope of 0.333, vs. 1/3 as predicted. It also gives $G_{cr} = 3 G_{avg}$, and the size distribution has a standard deviation $\sigma$ of 1 ($\sigma'=1/3$), which remains unchanged over $t$. So, all the analytical solutions are fully verified.

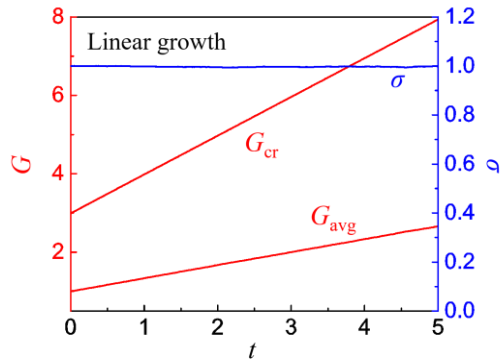

**Figure 5** Numerical calculations showing linear law for 3-grain line controlled growth. Initial grain distribution is from theoretically predicted one.

Next, we consider the case of $\alpha=2$ where 4-grain junctions control the growth, which has no predicted steady-state solution. Starting with a Gaussian size distribution, with a mean at 1 and a standard deviation of 0.1, and setting $\frac{2\gamma M_j}{a^2}=1$ plus a lower limit of the grain size at $G=a=0.001$ to avoid the singularity at $G=0$, we obtained the numerical results shown in **Fig. 6**. For comparison, **Fig. 6** also displays simulations for other growth laws using the same starting (non-steady-state) grain size distribution to examine the evolution toward their respective steady-states. For $\alpha=2$, while $G_{cr}$ indeed increases with time (**Fig. 6a**) and initially follows an exponential kinetics before it saturates (inset of **Fig. 6a**), $G_{avg}$ actually decreases with time (**Fig. 6b**). This is because grains of $G=0$ do not disappear given their $dG/dt=0$, so their number accumulates so much that they weigh down $G_{avg}$. Interestingly, there is an apparent transition in the growth kinetics at about $t=1.3$, marked by a dashed line in **Fig. 6**, which is accompanied by the emergence of very large grains and the saturation of $G_{cr}$ and $G_{max}$ in **Fig. 6a** and **d**. Numerically, the transition occurred when the population still contains a statistically significant number of grains (the inset of **Fig. 6d**)—a significant drop in population does not occur until $t=9$, after which the statistics becomes poor. Time sequence of grain size distribution in **Fig. 7** indicates the majority of the grains shrink rather than grow, shifting the grain size distribution towards left where small grains finally get stabilized at ~zero grain size. Meanwhile, a few large grains, with negligible portion in **Fig. 7**, grow uncontrollably at the expense of the shrinking grains, as can be seen from **Fig. 6d**. Such a bifurcation in grain size evolution is a key feature for 4-grain junction controlled growth. Furthermore, during this entire time, the standard deviation $\sigma$ (red curve in **Fig. 6c**) continues to increase, mostly with a concave upward shape, which confirms that the steady-state is unlikely to be approached. This is in contrast to all the other cases of smaller $\alpha$ whose $\sigma$ asymptotically approaches a steady-state value, and their $G_{avg}$ establishes their respective steady-state growth law relatively early in **Fig. 6a.** In fact, the smaller the $\alpha$, the faster the approach to the steady state. This trend reinforces our earlier observation that the higher the growth exponent, the less dispersive is the grain size distribution. Therefore, the failure to reach a steady-state kinetics and steady-state size distribution is limited to 4-grain junction control and any $1<\alpha\leq 2$.

In the above, we employed the mean-field approach by Lifshitz, Slyozov [1] and Hillert [3], which ignores topological features of the grain-grain boundary network. Parallelly, the well-known Von Neumann-Mullins relation [15,16] considers grain growth rate solely determined by the topological class of the grain—number of sides $x$

$$\frac{dS}{dt} = M_b \gamma \frac{\pi}{3}(x-6) \qquad (20)$$

where $S$ is the area of two-dimensional grain, while assuming identical grain boundary energies and mobilities independent of the grain size and equilibrium dihedral angles (120º) at triple grain junctions. The Von Neumann-Mullins relation has been re-visited under triple junction pinning [17,18], by assigning finite triple junction mobility and consequently shifting the dihedral angles from 120º. This approach is fundamentally identical to Lifshitz-Slyozov-Hillert approach if there exists a one-to-one correlation between the number of sides $x$ and the grain size $G$. Indeed, such a correlation has been sought by theory [19], simulation [20] and experiments [21], despite of a dispersion in the grain size within the same topological class $x$. Specifically, with the fitted correlation from Monte Carlo Potts model [20,22], the size distribution has been calculated with 3-grain line/4-grain junction pinning, which agrees well with the results given by the same

Potts-model simulations [23,24]. The calculated distributions have similar features as the ones in **Fig. 3** with a positive $\alpha$ close to one, which shows stability of small grains, none-zero $P'(u)$ and $P(G/G_{avg})$ at zero grain size, and an increased deviations $\sigma$ and $\sigma'$ in the relative grain sizes.

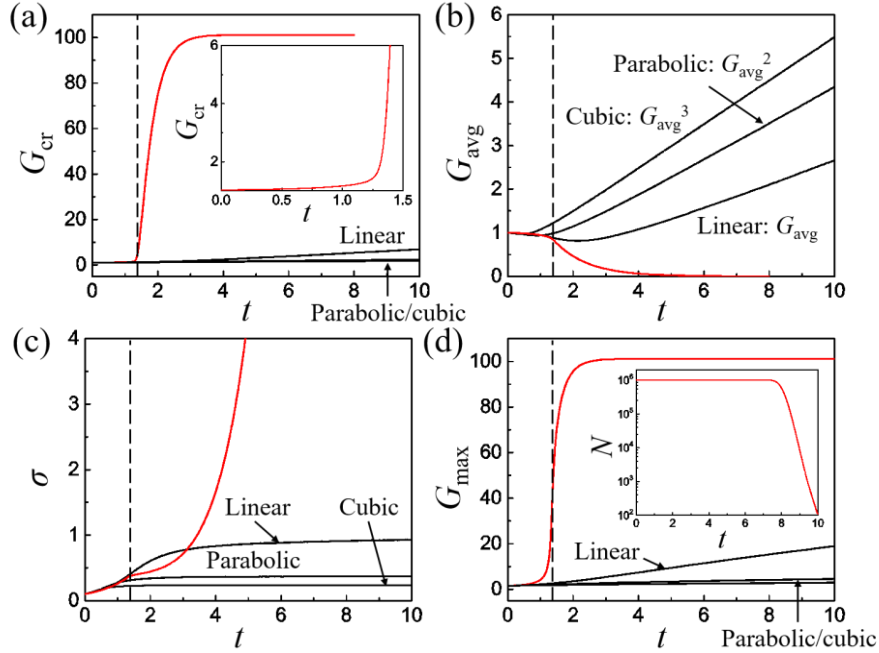

**Figure 6** Numerical results for 4-grain junction controlled growth, (a) critical grain size $G_{cr}$ (inset: $G_{cr}$ in the initial stage), (b) average grain size $G_{avg}$, (c) standard deviation $\sigma$ and (d) maximum grain size $G_{max}$ (inset: total number $N$ of the grains). The dash line indicates a transition where very large grains emerge and affect the overall growth kinetics. Also included as solid black curves are results leading to cubic, parabolic and linear growth laws. Same initial grain size distribution of a Gaussian one with mean at 1 and standard deviation of 0.1. Time $t$ is in the unit of $\left(\dfrac{2\gamma M_j}{a^2}\right)^{-1}$.

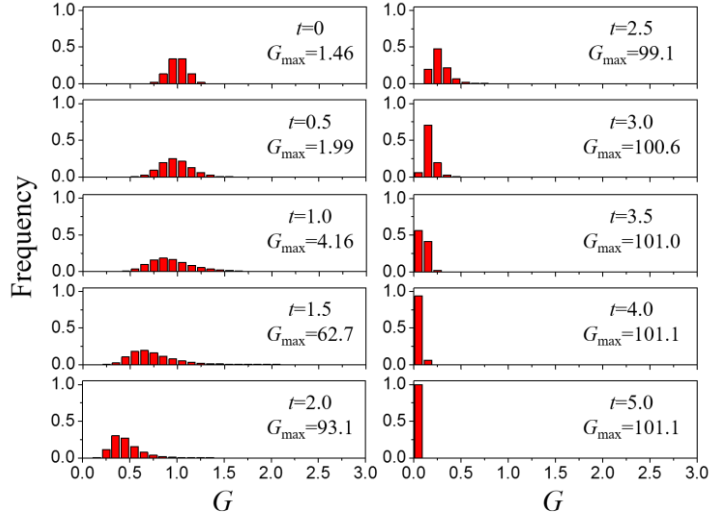

**Figure 7** Grain size distribution at different time for 4-grain junction controlled growth.

## III.  Growth Kinetics under Mixed Boundary/Junction Control

As mentioned in the Introduction, grain growth requires the motion of the entire grain boundary network, which includes 2-grain boundaries, 3-grain lines and 4-grain junctions. If one component controls the network motion, then the solution is already provided in **Section II**. If not, then it falls under mixed control, which may be solved by imposing the same velocity $v$ on all the components under their respective driving force, $F_b$ for a 2-grain boundary, $F_t$ for a 3-grain line and $F_j$ for a 4-grain junction

$$v = \frac{dG}{dt} = M_b F_b \qquad (21)$$

$$v = \frac{dG}{dt} = \frac{M_t G}{a} F_t \qquad (22)$$

$$v = \frac{dG}{dt} = \frac{M_j G^2}{a^2} F_j \qquad (23)$$

With the sum of $F_b$, $F_t$ and $F_j$ equal to the total capillary driving force

$$F = F_b + F_t + F_j = \left(\mu_0 - \frac{2\gamma}{G}\right) = 2\gamma\left(\frac{1}{G_{cr}} - \frac{1}{G}\right) \qquad (24)$$

we find the overall growth rate, equal to $v$, given by

$$\frac{dG}{dt} = 2\gamma \left( \frac{1}{M_b} + \frac{a}{M_t G} + \frac{a^2}{M_j G^2} \right)^{-1} \left( \frac{1}{G_{cr}} - \frac{1}{G} \right) \qquad (25)$$

Although this equation does not have an analytical solution, it can be numerically tackled in very much the same way as described above, with the following critical size $G_{cr}$

$$G_{cr} = \frac{\sum \left( \frac{1}{M_b} + \frac{a}{M_t G} + \frac{a^2}{M_j G^2} \right)^{-1} G^2}{\sum \left( \frac{1}{M_b} + \frac{a}{M_t G} + \frac{a^2}{M_j G^2} \right)^{-1} G} \qquad (26)$$

In our simulation, we started with the steady-state size distribution for 2-grain boundary control (Hillert's solution with $G_{avg}$=8/9), and set $2M_b\gamma$=1, $a$=0.001 and with various combinations of $M_t/M_b$ and $M_j/M_b$. (In the effective mobility $\left( \frac{1}{M_b} + \frac{a}{M_t G} + \frac{a^2}{M_j G^2} \right)^{-1}$, $G$ is much larger than $a$ so $M_t$ and $M_j$ need to be much smaller than $M_b$ in order to have a significant influence on the overall kinetics.) The numerical solutions obtained are described below.

The first set of simulation is presented in **Fig. 8**, where $M_j/M_b$=1 and $M_t/M_b$=1, $10^{-2}$, $10^{-3}$ and $10^{-4}$. As $M_t/M_b$ decreases, growth slows due to pinning by the 3-grain lines. Meanwhile, $\sigma$ increases, which is expected from **Fig. 6c** where $\sigma$ is larger for 3-grain line control. The growth kinetics deviate from the parabolic one in both the $G_{avg}$-$t$ plot in **Fig. 8a** (more evidently shown by $G_{avg}^2$-$t$ plot in **Fig. 8b**) and the $G_{cr}$-$t$ plot in Fig. 8c. At $M_t/M_b$=$10^{-4}$ (which makes it smaller than $a/G_{avg}$, which is ~5×$10^{-4}$), the 3-grain lines begin to take control, and a linear growth law can be identified after an initial transient. At higher $M_t/M_b$, however, the growth tends to return to parabolic growth after some initial slowdown.

The second set of simulations in presented in **Fig. 9**, where $M_t/M_b$=1 and $M_j/M_b$ = 1, $10^{-5}$,

$10^{-6}$ and $10^{-7}$. Again, as $M_j/M_b$ decreases, growth slows due to pinning by 4-grain junctions. Meanwhile, a larger $\sigma$ as expected emerges since there is no steady-state size distribution if grain growth is controlled by 4-grain junctions. The growth kinetics deviates from the parabolic one, and it becomes exponential in the case of very slow junction mobility ($G_{cr}$-$t$ plot in **Fig. 9d** at $M_j/M_b=10^{-7}$, which is smaller than $a^2/G_{avg}^2$), which is similar to the result obtained under solely 4-grain junction control in **Fig. 7**. At higher $M_j/M_b$, however, it is clear that the growth tends to return to parabolic growth after some initial slowdown, which confirms the same, though less pronounced trend seen in **Fig. 8**. This is understandable: As growth continues, the concentration of 4-grain junctions decreases, so the effect of a small $M_j/M_b$ diminishes with time. This effect is more less pronounced in **Fig. 8** because the concentration of 3-grain junctions decreases less rapidly than the 4-grain junctions.

We also simulated growth with other combinations of $M_t/M_b$ and $M_j/M_b$ and the observations were essentially the same. Summarizing these results, we may conclude the following. If $M_t/M_b<a/G_{avg}$ or $M_j/M_b<a^2/G_{avg}^2$, then there is significant growth slowdown by 3- or 4-grain junction pinning, which increases $\sigma$ and creates a more dispersed grain size distribution. Eventually, this also leads to a growth exponent $n$ smaller than 2, reaching $n=1$ under 3-grain-line control while the growth curve for $G_{cr}$ changes curvature becoming exponential under 4-grain-junction control.

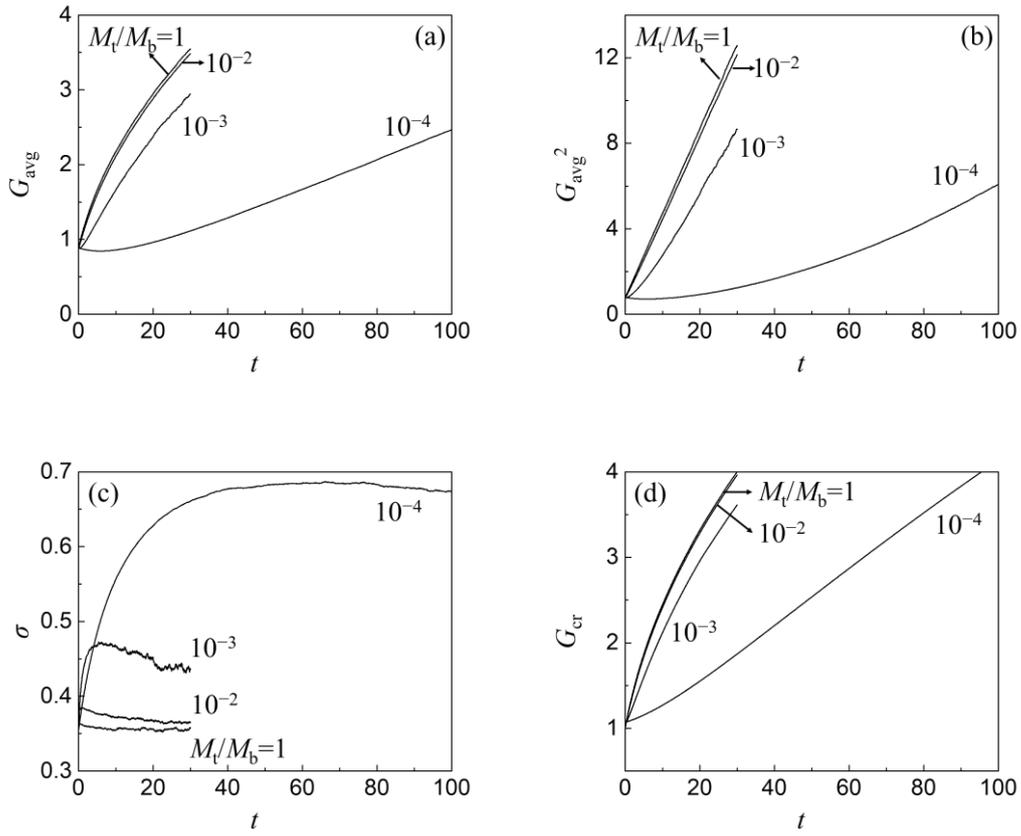

**Figure 8** Calculated (a) $G_{avg}$, (b) $G_{avg}^2$, (c) $\sigma$ and (d) $G_{cr}$ as a function of time $t$, with $M_j/M_b=1$ and different $M_t/M_b$.

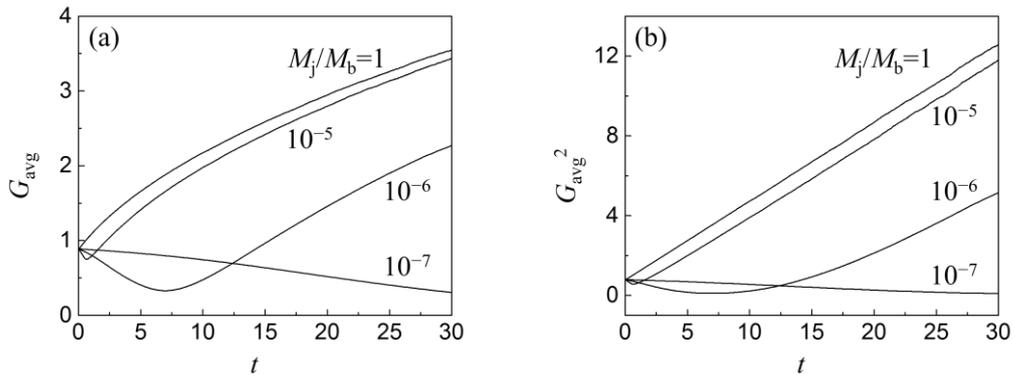

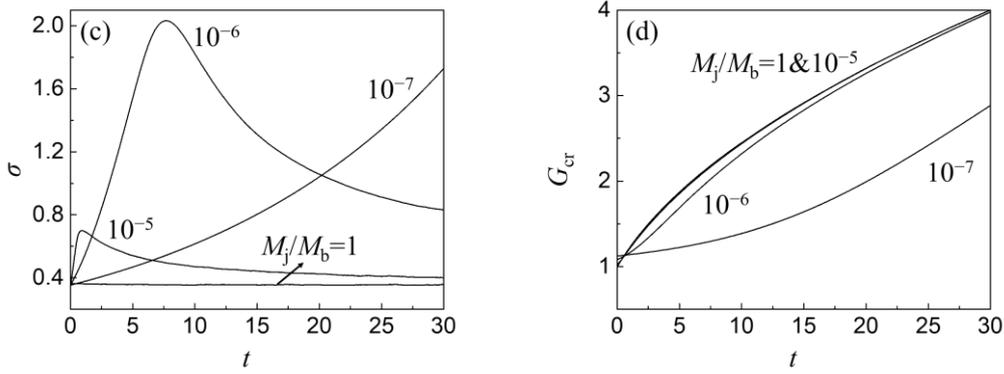

**Figure 9** Calculated (a) $G_{avg}$, (b) $G_{avg}^2$, (c) $\sigma$ and (d) $G_{cr}$ as a function of time $t$, with $M_t/M_b=1$ and different $M_j/M_b$.

## IV. Growth Kinetics with Statistically Distributed Mobility

The solutions given above can already provide insight to the effect of mobility inhomogeneities. Consider the case when $M_b$ decreases with grain size in a power law fashion, $M_b \sim G^{-\delta}$ with $\delta>0$, because of accumulation of solutes or second phase particles. This will cause $\alpha$ decreases from 0 to $-\delta$, which leads to an increase of the growth exponent $n$ and a decrease of $\sigma$ and $\sigma'$. This actually illustrates a general trend. In the mean-field theory, growth stagnation as reflected in a higher growth exponent is accompanied by a decrease in the standard deviation, because slowdown of the larger grains will allow smaller grains to catch up. Indeed, pinning by 3-grain lines and 4-grain junctions in mixed control growth illustrated in **Fig. 8-9** also follows the same trend. Below, we will examine whether such trend can be reversed by more severe inhomogeneities in mobilities, such as bimodal mobilities, resulting in both growth stagnation and increased size dispersion.

First, we examine the possibility of having different grain boundaries moving at different mobilities. (Our model includes the product of $M$ and $\gamma$, so the variation could also result from different grain boundary energies. But $\gamma$ is unlikely to vary by more than a factor of 3 or 4, whereas mobility is known to change by orders of magnitude.) In our first example, we let $\log(M_b\gamma)$ follow a Gaussian distribution with an average at $\log(M_{b,0}\gamma_0)$ and a standard deviation $\Sigma_b$. When $\Sigma_b=0$, it is mono-dispersed and all grain boundaries behave the same. Again, the size of each grain follows the equation of motion, Eq. (1), and the critical size $G_{cr}$ can now be calculated from

$$G_{cr} = \frac{\sum M_b \gamma G^2}{\sum M_b \gamma G} \qquad (27)$$

Starting with Hillert's size distribution (with $G_{avg}=8/9$) and setting $2M_{b,0}\gamma_0 = 1$, we varied $\Sigma_b$ to obtain the results shown in **Fig. 10a**. As $\Sigma_b$ increases, growth begins to slightly deviate from a parabolic one. More notable is the increase in $\sigma$ in **Fig. 10b**, which reaches about 0.45 at $\Sigma_b=0.08$ from the base line of 0.354. The latter change should be easily detectable experimentally (**Fig. 10** in Ref. 13).

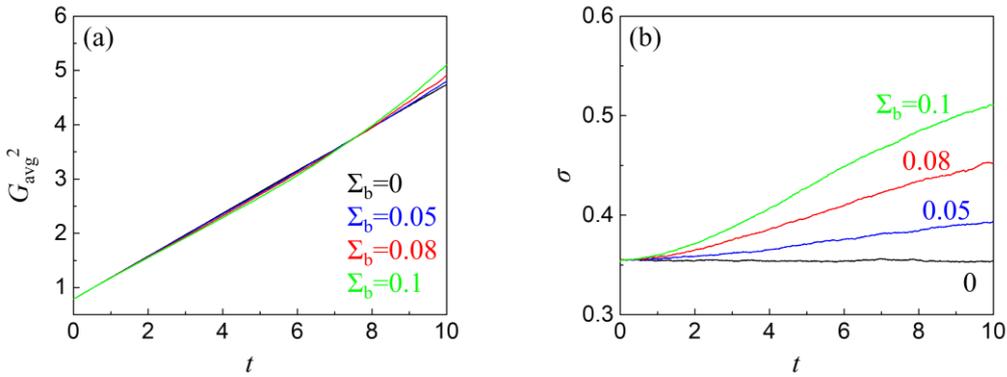

**Figure 10** Numerical calculations for 2-grain boundary controlled grain growth with statistically varied $M\gamma$, showing (a) $G_{avg}^2$, and (b) $\sigma$ as a function of time $t$ under different $\Sigma_b$.

The above model assumes that 2-grain boundaries are statistically inhomogeneous. From the size consideration, one may expect even more variation in $M_j\gamma$ and $M_t\gamma$ than in $M_b\gamma$. This is because $M_b\gamma$ is an average quantity over the entire 2-grain boundary of many atoms, while $M_t$ is only averaged over the 3-grain line with fewer atoms, and $M_j$ over a 4-grain junction with at most a few atoms. So the statistical variations are expected to follow $\Sigma_b<\Sigma_t<\Sigma_j$. Therefore, we next consider mixed control, varying $\Sigma_t$ for log ($M_t\gamma$) and $\Sigma_j$ for log ($M_j\gamma$) while keeping $\Sigma_b=008$ in all simulations. Taking $M_{j,0}/M_{b,0}=1$, $\Sigma_b=0.08$, $\Sigma_j=0$ and $2M_{b,0}\gamma_0=1$, under mixed 2-grain boundary and 3-grain line control, for example, we find as shown in **Fig. 11** a growth stagnation in the $G_{avg}^2$-$t$ plot accompanied by an increased $\sigma$ as $\Sigma_t$ is larger than 1.0. This is qualitatively a new feature not seen in all the analytic and numerical solutions thus far. A further decrease in $M_{t,0}/M_{b,0}$ will lead to an earlier onset of the above observations at a smaller $\Sigma_t$ (data not shown) but the same new feature remains. Interestingly, while in the case of $\Sigma_t=0$, the $G_{avg}^2$-$t$ plot in **Fig. 11** exhibits the feature of initial slowdown followed by the return of the parabolic growth, such feature disappears at larger dispersion, and the growth law more resembles the linear law than the parabolic law at longer time. This is understandable because at longer time, the grains with a lower $M_t$ are more likely to survive, while the ones with a higher $M_t$ shrink and disappear more rapidly [25]. Parallel calculations were also conducted with $M_{t,0}/M_{b,0}=1$, $\Sigma_b=0.08$, $\Sigma_t=0$ and $2M_{b,0}\gamma_0=1$, while varying $M_{j,0}/M_{b,0}$ and $\Sigma_j$ under mixed 2-grain boundary and 4-grain junction control. The same new feature is again confirmed, as shown in **Fig. 12**, as is the persistence of slowdown at longer time when $\Sigma_j$ is large. Therefore, by assigning the 3-grain lines and 4-grain junctions with their own mobilities and allowing statistical variations in boundary/junction mobilities/energies, we can obtain (i) smaller grains and decelerated grain growth with larger growth exponent $n$ from pinning for a prolonged time, and (ii) larger $\sigma$ hence more microstructural inhomogeneity. These features cannot be obtained from the solutions in **Section II** and **III** with uniform mobility, but they were seen in our experiments described in Ref. 13.

While the large $\Sigma_t$ and $\Sigma_j$ used in **Fig. 11** and **12** seem extreme as they imply up to 6 to 7 orders of magnitude difference in the mobilities of different 3-grain lines and 4-grain junctions, such large variations could simply be the result of a bimodal distribution. For example, there may be two sets of junctions, one mobile and the other immobile. [25] Therefore, while immobile junctions pin grain growth and increase $\sigma$ because of the size-dependent effective mobility $\frac{M_t G}{a}$, stagnation in $G_{\text{avg}}$ is gradually achieved when the number of (coarsening) grains with mobile junctions shrink, shifting the balance of the average toward the immobile, hence non-coarsening grains. This is demonstrated in **Fig. 13** in the following simulations. With respect to the reference case (black curves in **Fig. 13**) where all grain boundaries, lines and junctions are mobile and have identical mobilities $M_b=M_t=M_j=M$, we consider three special cases: (i) 10% of 2-grain boundaries in the initial population are immobile with a mobility of $M/10^4$, shown by the blue curves; (ii) 10% of 3-grain lines in the initial population are immobile with a mobility of $M/10^4$, shown by the red curves; (iii) 10% of 4-grain boundaries in the initial population are immobile with a mobility of $M/10^8$, shown by the green curves. The features are similar for the above three cases: grain growth slows down with a high growth exponent $n$, while microstructure becomes more inhomogeneous indicated by an increased $\sigma$, which agrees well with the trends in **Fig. 11** and **12**. Interestingly, when the heterogeneity in 2-grain boundary mobility is large enough (the mobility ratio between mobile and immobile 2-grain boundaries is $10^4$ in **Fig. 13**, much larger than the ones shown in **Fig. 10**), it also provides above trend. Therefore, we conclude the simultaneous growth stagnation and increased microstructural inhomogeneity can be directly explained by statistically distributed mobility of grain boundary/junction. However, the heterogeneity in the boundary/junction is required to be very large. Statistically, it suggests a more pronounced role of grain junctions than grain boundaries, since the former being averaged over much less atoms is expected to have larger variations.

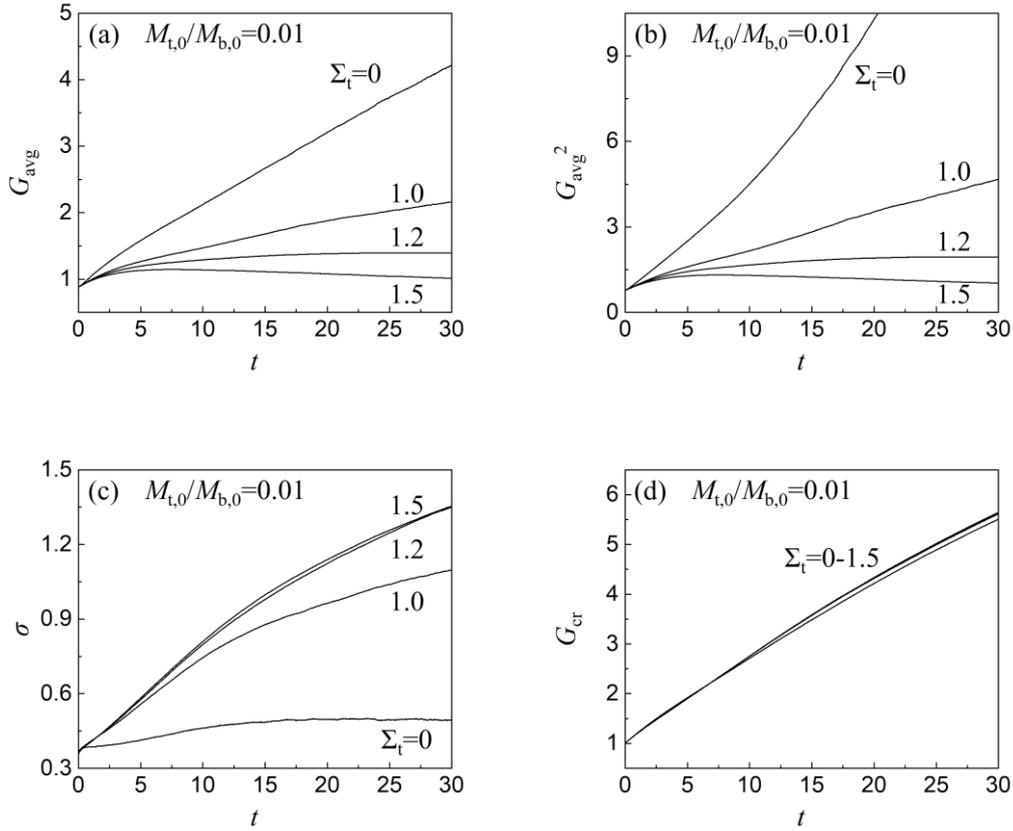

**Figure 11** Calculated (a) $G_{avg}$, (b) $G_{avg}^2$, (c) $\sigma$ and (d) $G_{cr}$ as a function of time $t$, with $M_{t,0}/M_{b,0}=0.01$, $M_{j,0}/M_{b,0}=1$, $\Sigma_b=0.08$, $\Sigma_j=0$ and different $\Sigma_t$.

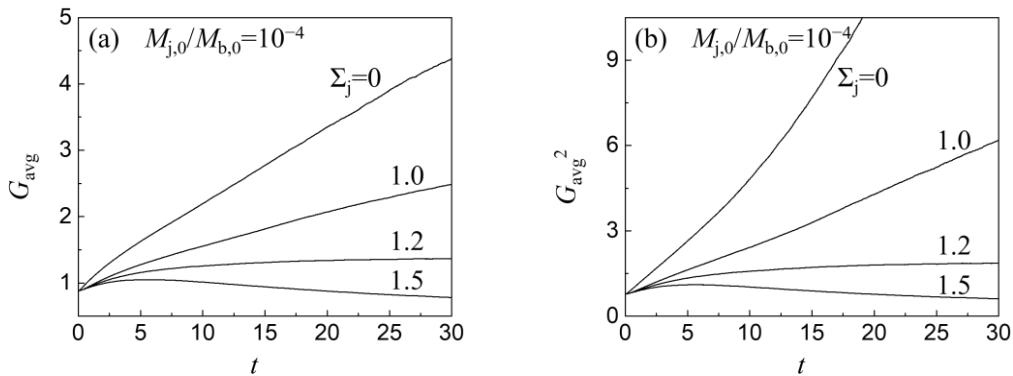

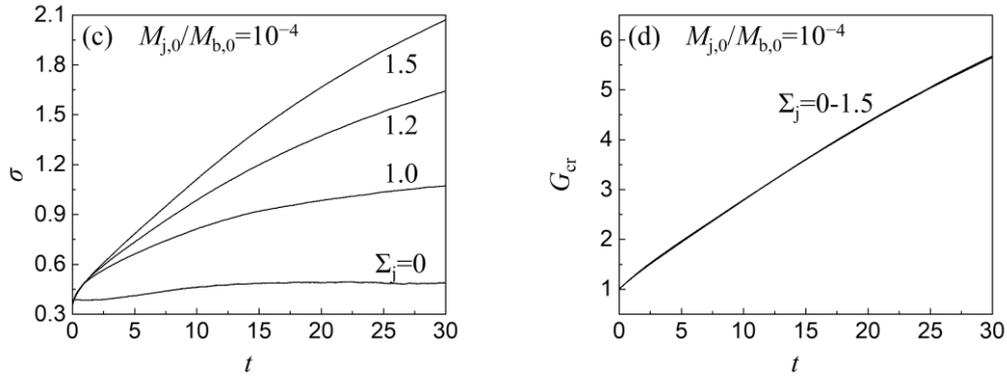

**Figure 12** Calculated (a) $G_{avg}$, (b) $G_{avg}^2$, (c) $\sigma$ and (d) $G_{cr}$ as a function of time $t$, with $M_{t,0}/M_{b,0}=1$, $M_{j,0}/M_{b,0}=10^{-4}$, $\Sigma_b=0.08$, $\Sigma_t=0$ and different $\Sigma_j$.

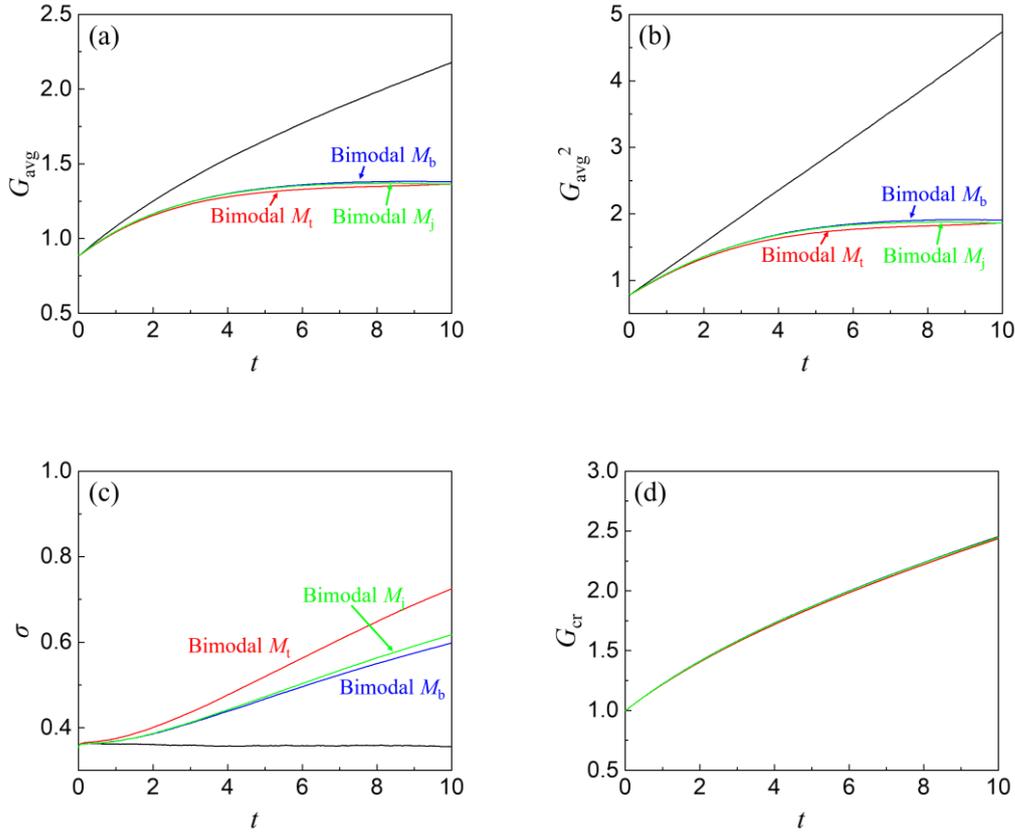

**Figure 13** Calculated (a) $G_{avg}$, (b) $G_{avg}^2$, (c) $\sigma$ and (d) $G_{cr}$ as a function of time $t$, with bimodal $M_b$ in blue, bimodal $M_t$ in red or bimodal $M_j$ in green. Reference curves in black are shown for the cases without bimodal mobilities. For mobile ones, we set $M_b=M_t=M_j=M$; for immobile ones, we set $M_b=M/10^4$, $M_t=M/10^4$ and $M_j=M/10^8$.

**V.     Conclusions**

(1) Analytic, steady-state solutions were obtained for a generalized growth problem with size-dependent mobility. The solution encompasses the Lifshitz, Slyozov, Wagner (LSW) and Hillert solutions in their respective special case, but also covers the case of junction-controlled growth, by either three-grain junctions or four-grain junctions.

(2) In the analytic solution, there is a one-to-one correspondence between $\alpha$ (the size dependence of the mobility), the growth exponent $n$, the upper cut-off grain size ($u_0$ and $G_{max}/G_{avg}$), and $\sigma$ (variation in the size distribution, i.e., structural homogeneity). As $\alpha$ increases, $n$ decreases, while $u_0$, $G_{max}/G_{avg}$ and $\sigma$ increases, indicating a less homogenous microstructure. Conversely, a smaller or even negative $\alpha$ decreases the growth rate of larger grains, causing $n$ to increase, growth to be self-limiting, and $\sigma$ to decrease, resulting in a more uniform microstructure.

(3) Growth under mixed 2-grain boundary, 3-grain line and 4-grain junction control was numerically simulated to reveal qualitatively similar features as the above, demonstrating junction pinning results in a smaller $n$ and a larger $\sigma$.

(4) Qualitatively different features of a larger $n$ and a larger $\sigma'$ can be obtained if statistical variations are introduced to the boundary/junction mobilities, which was experimentally observed at lower growth temperatures. When the variations in 2-grain boundary mobility/energy is small, its effect is subtler; it does not alter the growth exponent but it does produce a larger $\sigma$, which was also experimentally observed at lower growth temperatures. Therefore, one may conclude that variations in grain boundary/junction mobilities are common in real materials, especially at lower temperatures.

**Appendix Solution to steady-state grain size distribution**

For Case (I-II), we proceed to find their steady-state size distributions. With a known positive $A$, Eq. (9-9) yields

$$\frac{du}{d\tau} = \frac{A(u-1) - u^{2-\alpha}}{(2-\alpha)u^{1-\alpha}} \quad (A1)$$

At time $\tau$, denote the total number of grains by $N(\tau)$ and the number of grains between $u$ and $u+du$ is by $\varphi(u, \tau)du$, where the size distribution function is $\varphi(u, \tau)$, which satisfies the continuity equation in the grain-size space

$$\frac{\partial \varphi}{\partial \tau} + \frac{\partial}{\partial u}\left(\varphi \frac{du}{d\tau}\right) = 0 \quad (A2)$$

Introducing the trial solution

$$\varphi = \chi(\tau + \psi) \bigg/ \frac{du}{d\tau} \quad (A3)$$

where $\chi$ is a function of $\tau$ and $\psi$ and $\frac{du}{d\tau}$ is a function of $u$ only, we verify $\varphi$ can satisfy Eq. (9-15) if

$$\frac{d\psi}{du} = -1 \bigg/ \frac{du}{d\tau} \quad (A4)$$

The solution of $\psi$ is obtained by integration of Eq. (A4)

$$\psi = \int_0^u \frac{-(2-\alpha)u^{1-\alpha}}{A(u-1) - u^{2-\alpha}} du \quad (A5)$$

This integral can always be numerically evaluated and has a closed form solution when $\alpha$ is an integer. Next, the function $\chi$ can be calculated from the constraint that the total grain volume $K$ of a $\beta$-dimensional system ($\beta=2$ for 2D and 3 for 3D) is conserved

$$K = \int_0^{u_0} G^\beta \varphi \, du$$

$$= \int_0^{u_0} u^\beta G_{cr}^{\ \beta} \varphi \, du \qquad (A6)$$

$$= \int_0^{u_0} u^\beta \exp\left(\frac{\beta \tau}{2-\alpha}\right) \frac{\chi}{\frac{du}{d\tau}} du$$

Since the integral is independent of $\tau$ if and only if $\exp\left(\frac{\beta \tau}{2-\alpha}\right)\chi$ is independent of $\tau$, it is necessary for $\chi$ to be expressed as

$$\chi(\tau + \psi) = B \exp\left[-\frac{\beta}{2-\alpha}(\tau + \psi)\right] \qquad (A7)$$

where $B$ is a constant. So, from Eq. (A3),

$$\varphi = B \exp\left[-\frac{\beta}{2-\alpha}(\tau + \psi)\right] \Big/ \frac{du}{d\tau} \qquad (A8)$$

To determine $B$, we relate it to the total number of grains $N(\tau)$ by

$$\begin{aligned} N(\tau) &= \int_0^{u_0} \varphi \, du \\ &= B \exp\left(-\frac{\beta \tau}{2-\alpha}\right) \int_0^{u_0} \exp\left(-\frac{\beta \psi}{2-\alpha}\right) \Big/ \frac{du}{d\tau} \, du \\ &= -B \exp\left(-\frac{\beta \tau}{2-\alpha}\right) \int_0^{\infty} \exp\left(-\frac{\beta \psi}{2-\alpha}\right) d\psi \\ &= -\frac{2-\alpha}{\beta} B \exp\left(-\frac{\beta \tau}{2-\alpha}\right) \end{aligned} \qquad (A9)$$

Therefore, the normalized steady-state grain size distribution function $P'(u)$, defined as $\varphi / N$, is

$$\begin{aligned} P'(u) &= \frac{\varphi(u,\tau)}{N(\tau)} \\ &= -\frac{\beta}{2-\alpha} \exp\left(-\frac{\beta \psi}{2-\alpha}\right) \Big/ \frac{du}{d\tau} \\ &= \frac{\beta}{2-\alpha} \exp\left[-\frac{\beta}{2-\alpha} \int_0^u \frac{-(2-\alpha)u^{1-\alpha}}{A(u-1)-u^{2-\alpha}} du\right] \frac{(2-\alpha)u^{1-\alpha}}{u^{2-\alpha}-A(u-1)} \end{aligned} \qquad (A10)$$

Finally, the average grain size $u_{avg}$ and $G_{avg}$ can be calculated by

$$u_{avg} = \int_0^{u_0} u P'(u) \, du \qquad (A11)$$

$$G_{avg} = u_{avg} G_{cr} \qquad (A12)$$

Lastly, we recite the solutions of Lifshitz-Slyozov and Hillert, and compare them with the solution of $\alpha=1$. For Oswald ripening, the coarsening equation can be written in the more familiar term

$$\frac{dG}{dt} = \frac{2D\Omega\gamma}{G}\left(\frac{1}{G_{cr}} - \frac{1}{G}\right)\bigg/ k_B T \qquad (A13)$$

where $D$ is the diffusivity, $\Omega$ is the atomic volume and $k_B T$ has their usual meaning. Lifshitz and Slyozov gave the steady-state growth kinetics

$$G_{avg}^3 - G_0^3 = \left(\frac{4}{9}\right)(2D\Omega\gamma/k_B T)t \qquad (A14)$$

where $G_{cr}=G_{avg}$, and the steady-state size distribution $P'(u)$

$$P'(u) = 3^4 \cdot 2^{-\frac{5}{3}} \cdot e \cdot u^2 (u+3)^{-\frac{7}{3}} \left(\frac{3}{2} - u\right)^{-\frac{11}{3}} \exp\left(-\frac{1}{1-2u/3}\right) \qquad (A15)$$

with a standard deviation $\sigma=\sigma'=0.215$. For normal grain growth controlled by grain boundary mobility, Hillert gave the steady-state growth kinetics

$$G_{avg}^2 - G_0^2 = \left(\frac{32}{81}\right)(2M_b\gamma)t \qquad (A16)$$

where $G_{cr} = \frac{9}{8}G_{avg}$, and the steady-state size distribution $P'(u)$ is

$$P'(u) = \frac{3u \cdot (2e)^3}{(2-u)^5} \exp\left(\frac{-6}{2-u}\right) \qquad (A17)$$

with a standard deviation $\sigma'=0.314$ and $\sigma=0.354$. For $\alpha=1$ when 3-grain lines control grain growth, the predicted linear growth law is

$$G_{cr} - G_0 = \frac{2\gamma M_t}{a}t \qquad (A18)$$

where $G_{cr} = 3G_{avg}$, and the steady-state size distribution $P'(u)$ is

$$P'(u) = 3\exp(-3u) \qquad (A19)$$

with a standard deviation $\sigma'=1/3$ and $\sigma=1$.